\documentclass[aps,prl,reprint,floatfix]{revtex4-2}

\usepackage{amsmath}
\usepackage{amsfonts}
\usepackage{bm}
\usepackage{graphicx}

\usepackage{braket}
\usepackage{upgreek}
\usepackage{url}
\usepackage[hidelinks]{hyperref}

\graphicspath{{./}}

\begin{document}

\title{Stochastic Collision Theory of Magnetism in Radical Fluids}

\author{Yoshiaki Uchida}
\email{y.uchida.es@osaka-u.ac.jp}
\affiliation{Graduate School of Engineering Science, The University of Osaka, 1-3 Machikaneyama, Toyonaka, Osaka 560-8531, Japan}

\author{Ryohei Kishi}
\email{kishi.ryohei.es@osaka-u.ac.jp}
\affiliation{Graduate School of Engineering Science, The University of Osaka, 1-3 Machikaneyama, Toyonaka, Osaka 560-8531, Japan}

\begin{abstract}
How stochastic, microscopic events generate deterministic, macroscopic properties is a fundamental question in physics. We address this question by developing a quantum master equation model for concentrated radical solutions, in which stochastic molecular collisions govern the magnetic response of the system. The model implies that the first-order exchange contribution averages to zero over collisions, while the second-order term survives as an effective ferromagnetic coupling that enhances magnetization. The results are consistent with experimental trends in magnetic response that are not explained by conventional theories. The statistical-averaging origin of the mechanism suggests possible extensions to other soft-matter settings, including liquid crystals.
\end{abstract}

\maketitle

\section{Introduction}
A central question in statistical physics is how stochastic, microscopic dynamics generate deterministic macroscopic properties. Describing material properties in terms of spin degrees of freedom began with mean-field theory in magnetism; that framework was later carried over to the theory of liquid crystals. In that tradition, we address the question here for concentrated radical solutions, where stable radicals, paramagnetic molecules with non-metallic spin sources, are attracting attention as functional materials. Research is actively underway on applications such as controlling spins using light with stable radicals \cite{2025khariushin_Supramoleculardyadsphotogeneratedqubitcandidates,2019akita_Photomagneticeffectsmetalfreeliquidcrystals}, and developing stable radicals that emit light \cite{2019kato_LuminescentRadicalExcimerExcitedStateDynamicsLuminescentRadicalsDopedHostCrystals}. These materials are expected to serve as input/output devices in spin-based information processing. Spin relaxation, a factor often overlooked in attempts to realize magnetic ordering, is now gaining significant attention. The spin-lattice relaxation time in stable radicals with non-metallic elements as spin sources \cite{1976percival_Saturationrecoverymeasurementsspinlatticerelaxationtimesnitroxidessolution} is about three orders of magnitude longer than that of metal ions \cite{1961bloembergen_ProtonRelaxationTimesParamagneticSolutionsEffectsElectronSpinRelaxation}, suggesting superior information retention capabilities. Magnetic interactions during molecular collisions can induce spin polarization; nevertheless, this effect has traditionally been considered negligible \cite{2012molin_SpinExchangePrinciplesApplicationsChemistryBiology}.

Macroscopic magnetic properties in molecular materials are typically understood to arise from exchange interactions between stationary molecules in solids, which depend on the relative distances and orientations of adjacent molecules. Some stable radicals, for example, exhibit ferromagnetic transitions at cryogenic temperatures \cite{2025kuno_Synthesischaracterizationmagneticinteractionscarbazole9oxylitsderivatives}. Magnetic ordering requires exchange interactions strong enough to overcome thermal fluctuations \cite{1963mcconnell_FerromagnetismSolidFreeRadicals}. In conductive magnetic materials, localized spins form magnetic order due to the itinerant nature of electrons \cite{2010bonanni_storyhightemperatureferromagnetismsemiconductors}. In all these conventional cases, the magnetic properties are understood within a static or near-static framework, where molecular mobility and spin dynamics in fluid phases are not primary considerations.

\begin{figure}[t]
  \centering
  \includegraphics[width=\columnwidth]{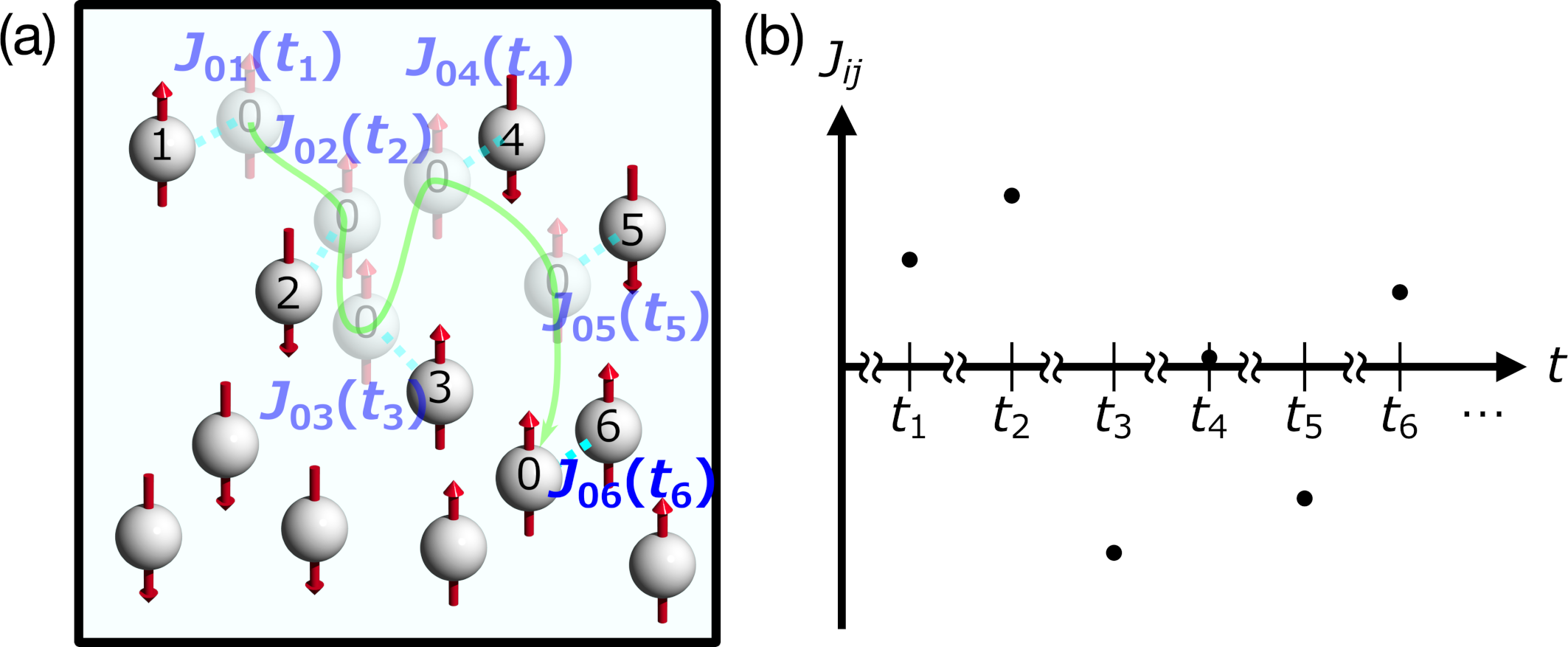}
  \caption{Stochastic exchange interactions arising from molecular collisions in a fluid.
(a) The model assumes that molecules undergo a series of discrete, pairwise collisions with varying partners.
(b) Each collision event generates an exchange interaction, $J_{ij}$, which is treated as a random variable fluctuating in both magnitude and sign (we take $J_{ij}$ to be Gaussian with zero mean in the model). This stochastic nature reflects the complex and unpredictable dynamics of molecular encounters in the fluid phase.}
  \label{fig:fig1}
\end{figure}

However, this conventional view is challenged by recent experimental findings. In high-concentration solutions of certain organic radicals, the magnetic susceptibility has been observed to increase at a solid-to-fluid phase transition, accompanied by a drastic increase in molecular mobility \cite{2008uchida_Unusualintermolecularmagneticinteractionobservedallorganicradicalliquidcrystal}. This phenomenon suggests that the magnetic susceptibility of the fluid phase is influenced by the dynamic magnetic interactions during molecular collisions \cite{2018nakagami_MolecularMobilityEffectMagneticInteractionsAllOrganicParamagneticLiquidCrystalNitroxideRadicalHydrogenBondingAcceptor}. The model uses a discrete-collision picture with stochastic exchange draws, as summarized in Fig.~\ref{fig:fig1}.

The key to understanding this unusual behavior lies in the competition between two timescales: the time interval between molecular collisions and the spin relaxation time (Supplementary Material Sec.~6). During a collision, magnetic interactions induce a perturbation in the spin polarization of the colliding molecules. After the collision, the induced spin polarization dissipates with time. Crucially, the spin-lattice relaxation time for the radicals is three to four orders of magnitude longer than the simulated time interval of the molecular collisions in the liquid-crystalline phase. This significant mismatch suggests that spin polarization induced by one collision may not fully relax before the next collision occurs, leading to an accumulation effect. While a previous model explained this by approximating molecular mobility as a contribution to the static coordination \cite{2020uchida_ThermalMolecularMotionCanAmplifyIntermolecularMagneticInteractions}, it remains unclear whether this approximation accurately represents the equilibrium state that arises from the time evolution of spin polarization.

Here, we directly address this question by simulating the time evolution of spin polarization under stochastic molecular collisions. We employ a quantum master equation model \cite{2002breuer_TheoryOpenQuantumSystems,2020manzano_ShortIntroductionLindbladMasterEquation} incorporating terms for both intermolecular interactions and detailed balance to compute the time evolution of the spin density matrix. Our calculations demonstrate that a sufficiently long relaxation time indeed leads to an increase in the magnetic susceptibility. The density matrix converges to a steady state, indicating that this phenomenon can be described as a form of ``itinerant molecular magnetism,'' in which the magnetic response is mediated by molecular motion rather than by a static lattice, arising from a dynamic equilibrium driven by molecular motion.

\section{Model}

We simulate the time evolution of the spin density matrix $\uprho$ under stochastic molecular collisions using a Lindblad-form quantum master equation \cite{1976lindblad_GeneratorsQuantumDynamicalSemigroups,1976gorini_CompletelyPositiveDynamicalSemigroupsNLevelSystems}, as shown in Fig.~1:
\begin{equation}
\frac{\mathrm{d}\uprho}{\mathrm{d}t} = -\frac{i}{\hbar} [\hat{H}_S, \uprho] + \mathcal{D}_{\mathrm{int}}(\uprho) + \mathcal{D}_{\mathrm{relax}}(\uprho),
\end{equation}
where $\mathcal{D}_{\mathrm{int}}$ and $\mathcal{D}_{\mathrm{relax}}$ account for collision-induced interactions and spin-lattice relaxation, respectively (Supplementary Material Secs.~1, 2, 7). We denote by $T_1$ the longitudinal spin--lattice relaxation time that controls relaxation of the diagonal populations toward the thermal state in the absence of exchange draws (Supplementary Material Sec.~2). The effective exchange interaction $J_{ij}$ during a collision fluctuates; we model it as a Gaussian random variable with zero mean \cite{1979palmer_Internalfielddistributionsmodelspinglasses,2020uchida_ThermalMolecularMotionCanAmplifyIntermolecularMagneticInteractions,1960kivelson_TheoryESRLinewidthsFreeRadicals}. This generates an effective magnetic flux density
\begin{equation}
B_{I}(t) = \frac{\sqrt{3}}{2} \frac{J_{ij}(t)}{g \upmu_{B}} + \frac{J_{ij}(t)^2}{\sqrt{3} (g \upmu_{B})^2 B}.
\end{equation}
Within this effective-field model, the first-order term in Eq.~(2), linear in $J_{ij}$, averages to zero over collisions because the exchange coupling fluctuates symmetrically about zero; the second-order term, $\propto J_{ij}^2$, is sign-definite and therefore provides the net contribution that enhances magnetization (Supplementary Material Secs.~3 and 4). The structure of the second-order bias follows the Thouless--Anderson--Palmer (TAP) mean-field framework for disordered exchange couplings \cite{1979palmer_Internalfielddistributionsmodelspinglasses}; \emph{TAP theory} supplies an effective single-spin field in which the sign-definite quadratic piece plays the role of a ferromagnetic bias. That second-order contribution is included following Palmer's cavity (``hole'') formulation of the effective field for the three-dimensional vector-spin case, as detailed for our collision-driven setting in Supplementary Material Sec.~3.

Compared with Redfield-type master equations \cite{1957redfield_OnTheoryRelaxationProcesses}, which are often employed when weak coupling to a harmonic bath sets temperature-dependent relaxation rates around a static reference Hamiltonian, the present approach treats each collision as an event that redraws $J_{ij}$ while retaining complete positivity through the Lindblad structure and enforcing lattice detailed balance through $\mathcal{D}_{\mathrm{relax}}$ (Supplementary Material Sec.~7). This event-based picture matches the concentrated-radical fluid setting emphasized here, where encounter rates and exchange magnitudes fluctuate strongly.

In simulations, this model is evaluated using the discrete-time, classical stochastic update rule summarized in Supplementary Material Sec.~8. Relaxation with time constant $T_1$ gradually counteracts this bias. From the steady-state populations we obtain the susceptibility
\begin{equation}
\upchi = \frac{C}{T} \left(1 + \frac{\Updelta B}{B}\right),
\end{equation}
where $\Updelta B$ is the interaction-induced shift in the magnetic flux density and $C$ is the Curie constant (Supplementary Material Sec.~5). The Results section compares the full and simplified models under identical conditions and then reports temperature and concentration dependences using the validated second-order-only formulation, which is computationally far more efficient.

\section{Results and discussion}
\subsection{Effect of First- and Second-Order Interaction Terms on Equilibration}
In this section, we numerically investigate the respective roles of the first- and second-order terms of the effective magnetic flux density $B_I(t)$ (Supplementary Material Sec.~3). We simulated the time evolution of the density matrix of the spin system under two distinct conditions: a ``full model'' incorporating both terms of $B_I(t)$, and a ``simplified model'' that includes only the second-order term. In all simulations, the interaction $J_{ij}$ was assumed to be a zero-mean random variable ($J_0=0$).

First, we conducted simulations using the full model. We used $T = 300$ K, $\upsigma_J = 1.1 \times 10^{-22}$ J, $\uptau_p = 1.00$ $\upmu$s, and $\uptau_{\mathrm{int}} = 0.22$ ns. These values are chosen to model concentrated organic radical solutions with frequent collisions, rather than the dilute-solution limit. Here $\uptau_{\mathrm{int}}$ is the mean interval between collision events (time step in the simulation), and $\uptau_p$ is the spin-lattice relaxation time; the hierarchy $\uptau_{\mathrm{int}} \ll \uptau_p$ allows spin polarization to accumulate (Supplementary Material Sec.~6). For the equilibration panels in Figs.~2(a) and 3(a), we use a magnetic flux density of $0.3\,\mathrm{T}$ (rather than $1\,\mathrm{T}$) so that the interacting steady populations are more clearly separated from the non-interacting Boltzmann baseline at the same field. The time evolution of the spin populations $\uprho_{00}$ and $\uprho_{11}$ exhibits large fluctuations in individual runs, but averaging over 1000 trials reveals a clear convergence to a steady state, as shown in Fig.~2(a). These fluctuations are a direct consequence of the first-order term of $B_I(t)$, which acts as a random scattering magnetic flux density. The equilibrated magnetization $M$ was then calculated and plotted against the magnetic flux density $B$, as shown in Fig.~2(b). We show $M(B)$ on two outer branches, $-1 \leq B \leq -0.3\,\mathrm{T}$ and $0.3 \leq B \leq 1\,\mathrm{T}$, where the effective-field expansion underlying Eq.~(2) is controlled and the TAP/Palmer bias is well defined. Following Palmer's cavity formulation of the effective internal field \cite{1979palmer_Internalfielddistributionsmodelspinglasses}, this effective-field description assumes a sufficiently large external magnetic field; Supplementary Material Sec.~3 records the same requirement explicitly by stipulating that the linear Zeeman splitting dominates when using the closed form for $B_I(B)$. We omit the central interval where $|B| < 0.3\,\mathrm{T}$ (approximately $-0.3 < B < 0.3\,\mathrm{T}$): there $|B|$ is not large compared with interaction-induced scales retained in the truncation, the parametrisation is inappropriate, and the raw magnetisation curves show singular small-$|B|$ behaviour that is an artefact of the strong-field truncation rather than a controlled prediction of the weak-field physics. Extending the analysis to that regime would require additional terms and/or revisiting the Palmer/TAP effective-field framework; we do not pursue it here. This result confirms that $|M| > |M_0|$, i.e., the magnetization is enhanced compared to that of a non-interacting system ($M_0$).

\begin{figure}[!t]
  \centering
  \includegraphics[width=0.55\columnwidth]{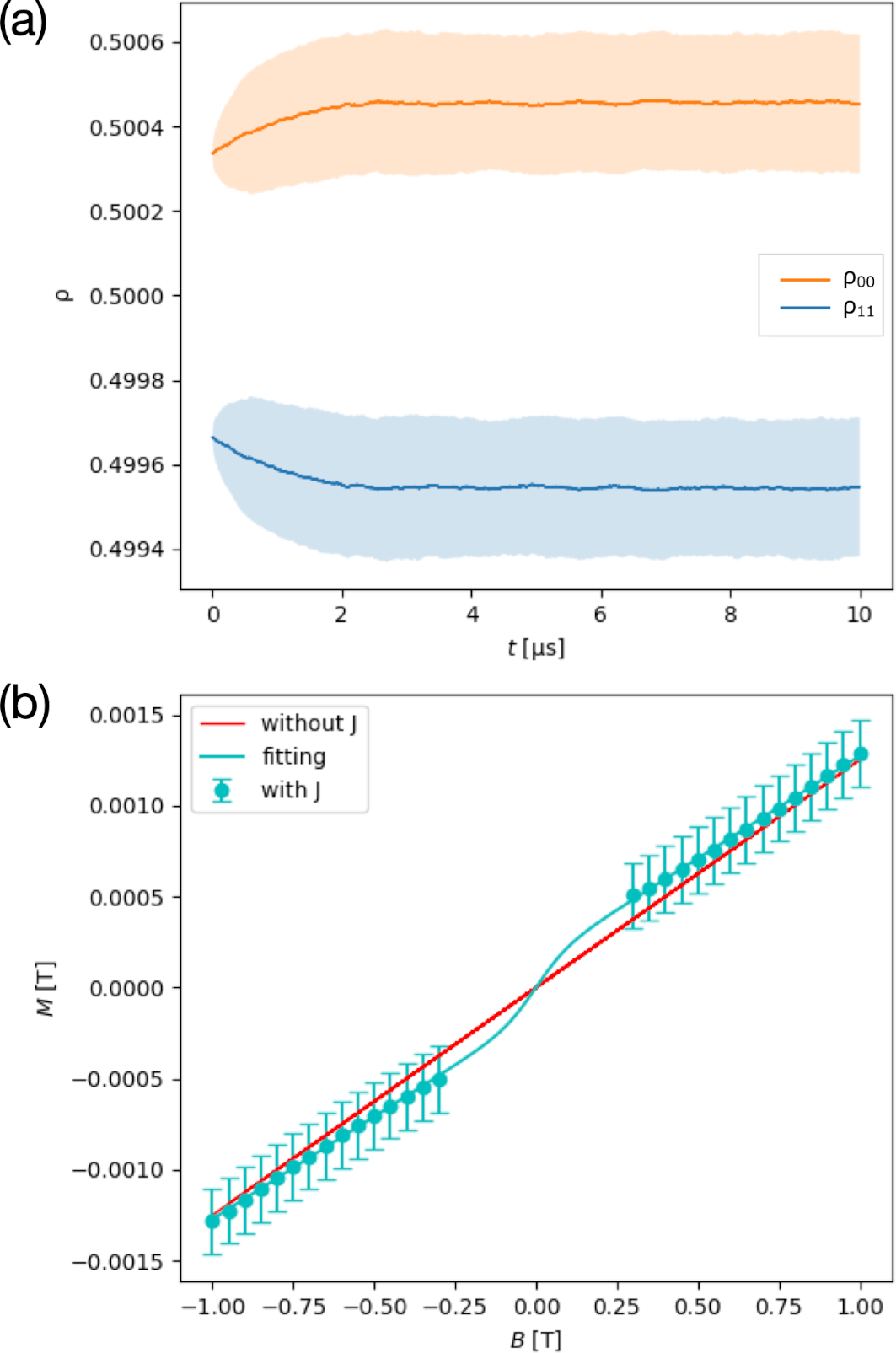}
  \caption{Effect of the first-order term (full model) on spin equilibration and magnetization.
(a) Time evolution of the spin populations $\uprho_{00}$ (orange) and $\uprho_{11}$ (blue) toward equilibrium at a fixed magnetic flux density of $0.3\,\mathrm{T}$. (b) The resulting magnetization curve $M$ (cyan line) satisfies $|M| > |M_0|$ compared to the ideal paramagnetic magnetization $M_0$ (red line), indicating enhancement from the interaction. $M(B)$ is plotted on two outer $|B|$ ranges only ($-1$--$-0.3$ and $0.3$--$1\,\mathrm{T}$); the central near-zero-$|B|$ interval is omitted because the truncation is unreliable there (see text). Shaded area in (a) denotes the standard deviation.}
  \label{fig:fig2}
\end{figure}

\begin{figure}[!t]
  \centering
  \includegraphics[width=0.55\columnwidth]{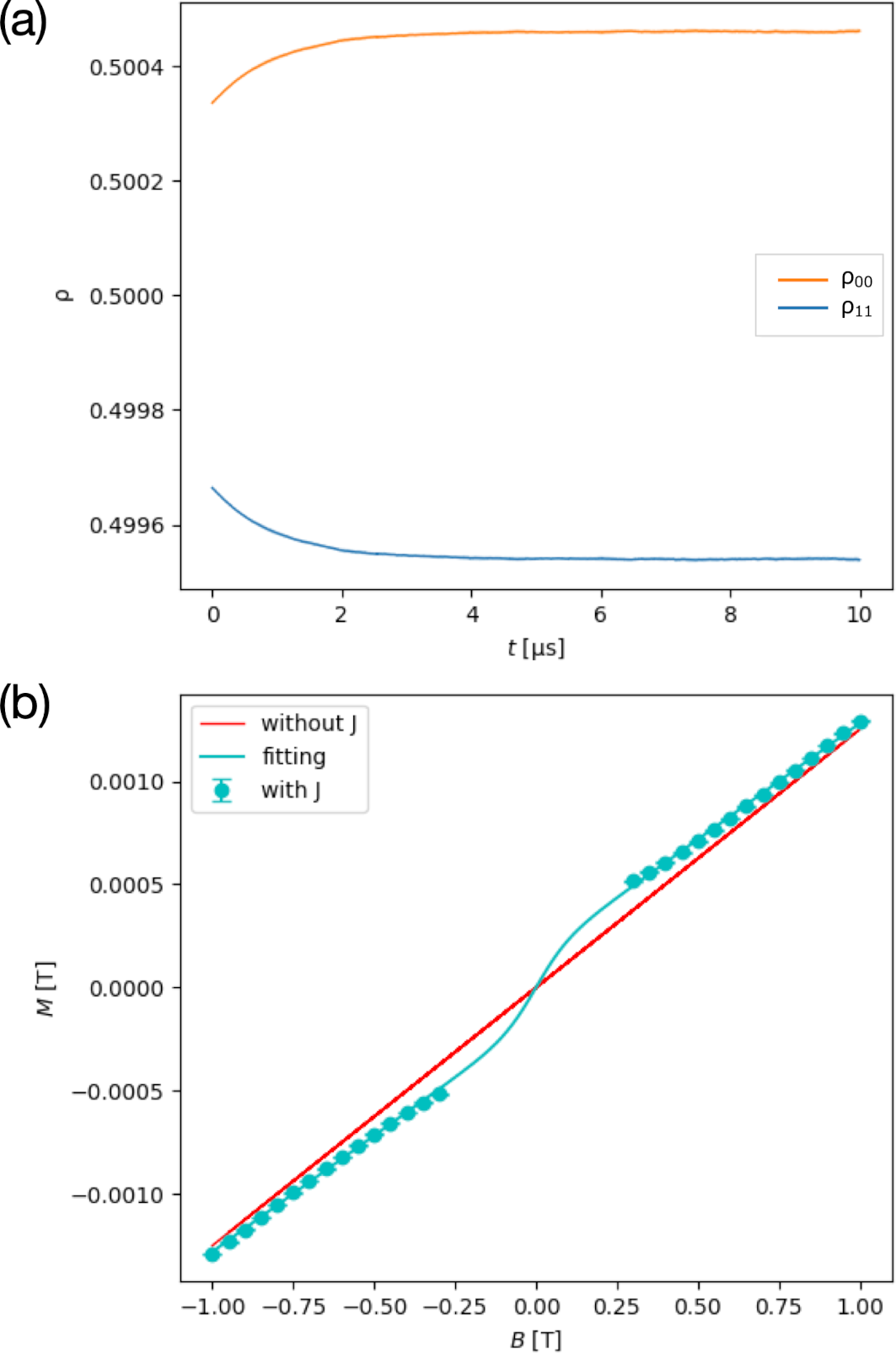}
  \caption{Effect of the second-order term (simplified model) on spin equilibration and magnetization.
(a) Time evolution of the spin populations $\uprho_{00}$ (orange) and $\uprho_{11}$ (blue) toward equilibrium at a fixed magnetic flux density of $0.3\,\mathrm{T}$. (b) The resulting magnetization curve $M$ (cyan line) satisfies $|M| > |M_0|$ compared to the ideal paramagnetic magnetization $M_0$ (red line), demonstrating the significant contribution of the second-order term. As in Fig.~2(b), only the two outer $|B|$ ranges are shown and the central interval is omitted (see text). Shaded area in (a) denotes the standard deviation.}
  \label{fig:fig3}
\end{figure}

Next, to isolate the contribution of the second-order term, we performed simulations using the simplified model. The system converges to equilibrium much more rapidly and monotonically, as shown in Fig.~3(a). Only 10 trials were needed to obtain a smooth curve. The resulting magnetization is enhanced compared to $M_0$, as shown in Fig.~3(b); as in the full model (Fig.~2(b)), $|M| > |M_0|$, and Fig.~3(b) uses the same two-branch $|B|$ coverage and the same omission of the central weak-field interval. This efficient convergence arises because the simplified model omits the first-order random-scattering term and retains only the sign-definite second-order term ($\propto J_{ij}^2$), which drives the system toward equilibrium.

The most crucial insight is gained by comparing the final equilibrium states of both models. The population values reached with the simplified model are nearly identical to those obtained from the computationally demanding full model, as shown in Figs.~2(a) and 3(a). This provides compelling numerical evidence that when the average interaction is zero, the first-order term in $B_I(t)$ does not affect the macroscopic equilibrium state, although it dominates the transient dynamics. This finding validates the physical intuition described in the Model section from a dynamic simulation perspective and supports the hypothesis that molecular mobility enhances the magnetic response through the second-order exchange term \cite{2020uchida_ThermalMolecularMotionCanAmplifyIntermolecularMagneticInteractions}. Furthermore, since the simplified model accurately reproduces the equilibrium properties with significantly reduced computational cost, it serves as both a sufficient and efficient tool for analysis. In the same simplified model, we also evolved the same $(T,B)=(300\,\mathrm{K},0.3\,\mathrm{T})$ conditions starting from extremal diagonal states $(\uprho_{00},\uprho_{11})=(1,0)$ and $(0,1)$ instead of the Boltzmann equilibrium used in Figs.~2 and 3; both runs approach the same steady-state magnetization as the Boltzmann-initialized trajectories within stochastic uncertainty (Supplementary Material Fig.~S1), which supports that the long-time magnetization is not tied to the particular near-thermal seed chosen for the main panels. Therefore, we will employ this validated, simplified model for all subsequent investigations of temperature and concentration dependence.

\subsection{Temperature Dependence}
\label{sec:temp_dep}

\begin{figure}[!t]
  \centering
  \includegraphics[width=0.55\columnwidth]{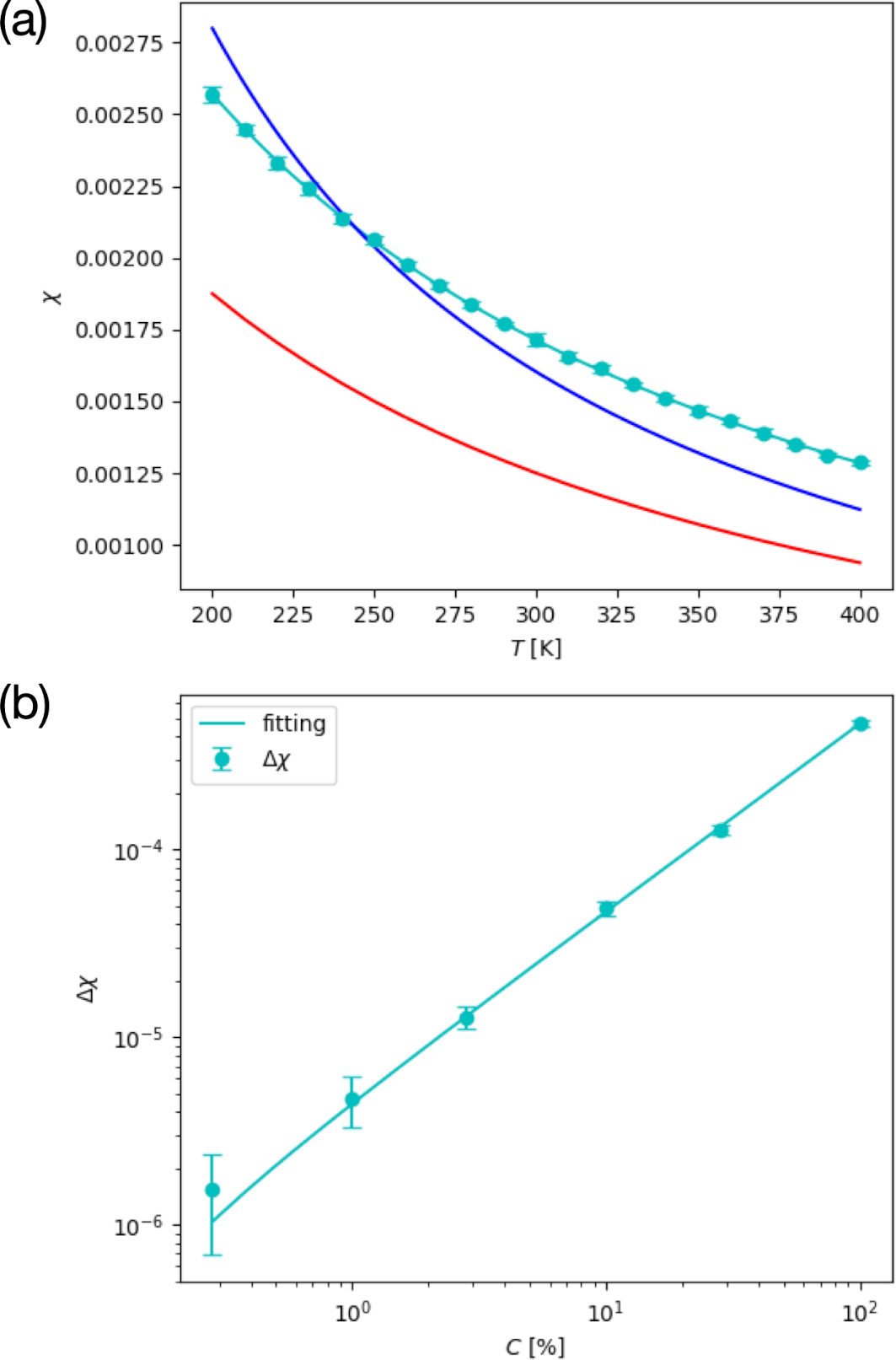}
  \caption{Temperature and concentration dependence of magnetic susceptibility.
(a) Magnetic susceptibility $\upchi$ as a function of temperature under a magnetic flux density of 0.3 T; the horizontal axis spans $200$--$400\,\mathrm{K}$ as in the experimental comparison (vertical axis: $\upchi$). The cyan circles represent the simulation data. The red solid line shows the susceptibility of non-interacting spins ($\upchi_0$). The blue solid line represents the best fit using the standard Curie-Weiss law ($\uptheta$ = constant), while the cyan solid line shows the fit with the modified Curie-Weiss law incorporating a temperature-dependent Weiss constant ($\uptheta(T) = aT+b$).
(b) Concentration dependence at 300 K and 0.3 T (vertical axis: $\Updelta\upchi$). The quantity is plotted against radical concentration (weight percent). A linear fit $\Updelta\upchi = a C + b$ to the simulation points yields $a \approx 4.68\times 10^{-6}$ per \% and $b \approx -2.8\times 10^{-7}$.}
  \label{fig:fig4}
\end{figure}

Based on the findings in the previous section, we proceed with our analysis using the simplified model, which considers only the second-order interaction term. We first investigated the temperature dependence of the magnetic susceptibility, as the anomalous thermomagnetic effect is typically observed between 200 K and 400 K. The calculations were performed at 10 K intervals within this range under a magnetic flux density of 0.3 T, a typical condition for X-band EPR spectroscopy.

The calculated magnetic susceptibility decreases with increasing temperature, as shown in Fig.~4(a). While this behavior qualitatively resembles the Curie-Weiss law, a quantitative analysis reveals a notable deviation. We attempted to fit the simulation data with the standard Curie-Weiss equation:
\begin{equation}
\upchi = \frac{C}{T - \uptheta},
\end{equation}
where the Curie constant $C$ was fixed at 0.375 emu K mol$^{-1}$ (for S=1/2) and $\uptheta$ is the Weiss constant. The best-fit constant $\uptheta \approx 66\,\mathrm{K}$ (formal uncertainty of order $3\,\mathrm{K}$) still leaves a systematic mismatch with the simulation points, as shown in Fig.~4(a).

This discrepancy suggests that the Weiss constant $\uptheta$ is temperature-dependent in our model. In solid-state systems, $\uptheta$ is correlated with the strength of intermolecular interactions within a fixed crystal structure and is thus generally temperature-independent. In contrast, for molecules in a fluid, the effective interaction strength is inherently dependent on thermal motion and collision dynamics, which vary with temperature. This dependence is the primary reason for the failure of the standard Curie-Weiss law to describe our system.

Indeed, similar deviations have been reported in previous experimental studies \cite{2008uchida_Unusualintermolecularmagneticinteractionobservedallorganicradicalliquidcrystal,2010uchida_AnisotropicInhomogeneousMagneticInteractionsObservedAllOrganicNitroxideRadicalLiquidCrystals}. Therefore, we introduced a linear temperature dependence for the Weiss constant, $\uptheta(T) = aT + b$. This modified model provided an excellent fit to our data with $\uptheta(T) \approx 0.270\,T + 0.074$ (in kelvin), as shown in Fig.~4(a). At absolute zero, $\uptheta(0) = b \approx 0.07\,\mathrm{K}$ is of the same order as the Weiss constant in typical organic radical crystals, which supports the credibility of the fit. The positive coefficient of the linear term in $T$ implies that the more vigorous the molecular motion, the stronger the effect, consistent with the hypothesis. This result demonstrates that our simulation effectively captures the characteristics of spin interactions in a fluid. It also suggests that employing a temperature-dependent Weiss constant is crucial for accurately analyzing experimental data in such systems.

\subsection{Concentration Dependence}
\label{sec:conc_dep}

To further explore experimentally verifiable predictions, we examined the concentration dependence of the equilibrium magnetization. A decrease in the concentration of organic radicals in a diamagnetic solvent should lead to a lower collision frequency among the radicals. Our time-evolution model is sensitive to this collision frequency. To model this, we assume that the collision frequency is directly proportional to the radical concentration \cite{1960kivelson_TheoryESRLinewidthsFreeRadicals}. This assumption relies on the simplification that the spin-carrying molecules and the solvent molecules have similar sizes and that the density and viscosity of the solution do not change significantly upon mixing.

Using as a reference the previously reported 100$\%$ radical liquid-crystal case \cite{2020uchida_ThermalMolecularMotionCanAmplifyIntermolecularMagneticInteractions}, we performed additional simulations at six concentrations from 100$\%$ down to approximately 0.28$\%$, with logarithmically spaced values of the collision frequency. We define the excess susceptibility $\Updelta\upchi = (\upchi - \upchi_0)/\upchi_0$, where $\upchi_0$ is the susceptibility of non-interacting spins. The calculated $\Updelta\upchi$ at 300 K and 0.3 T is plotted as a function of concentration, as shown in Fig.~4(b). Over the simulated range, $\Updelta\upchi$ is well described by a linear dependence on weight percent, $\Updelta\upchi \approx a C + b$ with $a \approx 4.68\times 10^{-6}$ per \% and a small offset $b \approx -2.8\times 10^{-7}$. The results indicate that the magnetic susceptibility per radical decreases as the concentration is reduced. This is an expected consequence, as lower concentrations lead to weaker effective intermolecular interactions, causing the behavior of the system to approach that of isolated, non-interacting spins, for which interaction effects are absent. This trend is consistent with the common experimental practice of studying radicals at low concentration when aiming to access intrinsic single-molecule properties, where intermolecular effects are minimized. This monotonic dependence on concentration provides a clear and experimentally testable prediction of our model. This prediction can be tested in existing experimental systems, such as nitroxide radical--diamagnetic liquid crystal mixtures, by varying composition and phase.

\section{Conclusion}

The main result of this work is that stochastic collisions in concentrated radical solutions yield an effective ferromagnetic coupling from the second-order exchange term, which explains the anomalous magnetic susceptibility observed in such fluids. We have demonstrated this by simulating the time evolution of the spin density matrix under a Lindblad master equation and by showing that the first-order term averages to zero over collisions while the second-order term survives. We have reproduced the anomalous temperature dependence observed in earlier experiments and proposed a testable concentration dependence for future verification. Our results support the hypothesis that molecular mobility enhances the magnetic response through the second-order exchange term \cite{2020uchida_ThermalMolecularMotionCanAmplifyIntermolecularMagneticInteractions}. This finding provides a consistent explanation for experimental results and offers a new paradigm for understanding how processes that drive the system toward equilibrium can dictate the resulting equilibrium properties. A controlled treatment of magnetisation in the very weak field regime, where the present TAP/Palmer truncation is not applicable, is left to future work.

The true significance of our approach lies in its generality. In the same spirit as the tradition in which a spin-based view of material properties began with mean-field theory in magnetism and was later carried over to liquid crystals, the mechanism unveiled here---a macroscopic property emerging from the statistical residue of stochastic microscopic interactions---is not unique to spin systems. We propose that this theoretical framework provides a unified foundation for investigating a broader class of phenomena in soft matter and chemical physics. In particular, the present picture---where the first-order fluctuating term does not control the final equilibrium and a reduced two-state statistical description captures the dominant effect---is consistent with classical orientational-transition modeling for flexible molecules \cite{1979kimura_OrientationalPhaseTransitionFlexibleMolecules}, and suggests useful links to shape-controlled chirality transfer in nematics \cite{1995ferrarini_ShapeModelTwistingPower} and recent molecular-level perspectives on liquid-crystal chirality \cite{2026yoshida_ChiralityLiquidCrystalsReview}. This work, therefore, opens a new avenue for exploring and predicting the collective properties of complex fluids governed by microscopic dynamics.

\section*{Acknowledgments}
This work was supported by JSPS KAKENHI Grant Numbers JP22H02158 and JP23K23426. The computation was performed using the Research Center for Computational Science, Okazaki, Japan (Project: 24-IMS-C050, 25-IMS-C051). We thank Dr. Hajime Miyamoto for reading the manuscript and for his helpful comments on the theoretical aspects of this work.

\bibliographystyle{apsrev4-1}
\bibliography{Manuscript_refs}

\end{document}